\documentclass[journal,twocolumn]{IEEEtran}
\usepackage{cite}
\usepackage{amsmath,amsfonts,amsxtra,amssymb,latexsym,amscd,amsthm,mathrsfs,bm}
\usepackage{tabularx}
\usepackage{graphicx}
\usepackage{hyperref}
\usepackage{xcolor}
\usepackage{algorithm}
\usepackage{algorithmic}
\usepackage{cite}
\usepackage[colorinlistoftodos,textsize=tiny]{todonotes}
\usepackage{balance}
\usepackage{epstopdf}
\usepackage{balance}
\usepackage{booktabs}

\usepackage{caption}
\usepackage{subcaption}

\newcommand{\dd}{{\mathrm d}}
\usepackage{eqparbox}

\usepackage{etoolbox}  % patch def of algorithmic environment
\makeatletter
\patchcmd{\algorithmic}{\addtolength{\ALC@tlm}{\leftmargin} }{\addtolength{\ALC@tlm}{\leftmargin}}{}{}
\makeatother

\title{Concurrent Transmission and Multiuser Detection of LoRa Signals
	\thanks{The authors are with the Department of Electrical and Computer Engineering, University of Saskatchewan, Saskatoon, Canada S7N5A9. Emails: \{khai.nguyen, ha.nguyen, e.bedeer\}@usask.ca.}
}

\author{\IEEEauthorblockN{The Khai Nguyen, Ha H. Nguyen, and Ebrahim Bedeer}}
\begin{document}
\maketitle

\begin{abstract}
This paper investigates a new model to improve the scalability of low-power long-range (LoRa) networks by allowing multiple end devices (EDs) to simultaneously communicate with multiple multi-antenna gateways on the same frequency band and using the same spreading factor. The maximum likelihood (ML) decision rule is first derived for non-coherent detection of information bits transmitted by multiple devices. To overcome the high complexity of the ML detection, we propose a sub-optimal two-stage detection algorithm to balance the computational complexity and error performance. In the first stage, we identify transmit chirps (without knowing which EDs transmit them). In the second stage, we determine the EDs that transmit the specific chirps identified from the first stage. To improve the detection performance in the second stage, we also optimize the transmit powers of EDs to minimize the similarity, measured by the Jaccard coefficient, between the received powers of any pair of EDs. As the power control optimization problem is non-convex, we use concepts from successive convex approximation to transform it to an approximate convex optimization problem that can be solved iteratively and guaranteed to reach a sub-optimal solution. Simulation results demonstrate and justify the tradeoff between transmit power penalties and network scalability of the proposed LoRa network model. In particular, by allowing concurrent transmission of 2 or 3 EDs, the uplink capacity of the proposed network can be doubled or tripled over that of a conventional LoRa network, albeit at the expense of additional 3.0 or 4.7 dB transmit power.
\end{abstract}


\section{Introduction}
Technical advances and applications in the Internet of Things (IoT) domain continue to evolve in recent years in order to support communications and connectivity of billions of end devices (EDs) worldwide \cite{figueredo2020preparing}. In many IoT applications, EDs need to communicate over distances of tens of kilometers with very low power consumption while being served by a few gateways (GWs). To satisfy such large coverage and low power consumption requirements, low-power wide-area networks (LPWANs) have been designed and deployed. Low-power long-range (LoRa) is one of the leading LPWAN technologies and is based on chirp spread spectrum (CSS), commonly refereed to in the literature as LoRa modulation in the PHY layer, and LoRaWAN protocol in the MAC layer \cite{centenaro2016long}.

In the PHY layer, LoRa modulation can be configured with three different bandwidths of 125 kHz, 250 kHz, and 500 kHz, as well as six different spreading factor (SF) values, from 7 to 12. Using a higher spreading factor increases the coverage range, but at the expense of a lower data rate \cite{afisiadis2019error}. In the MAC layer, the LoRaWAN protocol adopts pure ALOHA due to its simplicity and little communication overhead. However, pure ALOHA has its own shortcomings, the most critical of which is that only a limited number of EDs can access the channel at a given time, which reduces the scalability of LoRa networks~\cite{mahmood2018scalability}.

In the uplink transmission of a typical LoRa network, single-antenna EDs communicate with a number of single-antenna GWs. Then, GWs forward the received packets to the LoRa network server (LNS) along with the received signal strength indicator (RSSI) of each packet and optional time stamps. In the downlink transmission, the LNS communicates with a given ED through the GW with the highest RSSI.

Most research works on LoRa modulation are concerned with communication between a single ED and multiple GWs, where both the ED and GWs are equipped with a single antenna \cite{hanif2020US,hanif2021frequency, elshabrawy2018closed, hanif2020slope, nguyen2019efficient}. For example, in \cite{elshabrawy2018closed}, the authors derived tight closed-form approximations for the bit error rate (BER) of the conventional LoRa modulation in both additive white Gaussian noise (AWGN) and Rayleigh fading channels. In \cite{hanif2020slope}, the authors presented a method to increase data rates of the conventional LoRa modulation by adding a down chirp and its cyclic shifts to the signal set. Such a scheme is called slope-shift keying LoRa (SSK-LoRa). The authors also derived tight approximations for the BER and symbol error rate (SER) for non-coherent detection in Rayleigh fading channels. Another more flexible and advanced scheme is proposed in \cite{hanif2020US,hanif2021frequency} and called frequency-shift chirp spread spectrum with index modulation (FSCSS-IM), which can offer much higher data rates than the conventional LoRa modulation. The higher data rates of FSCSS-IM are achieved without the need to increase the transmission bandwidth, but at the cost of a slight deterioration of the BER. The authors also proposed a low-complexity near-optimal non-coherent detection algorithm whose performance approach its counterpart of the coherent detection. More recently, by extending the framework of FSCSS-IM in \cite{hanif2020US,hanif2021frequency} to the quadrature dimension, the authors in \cite{baruffa2021} proposed a scheme called  quadrature chirp index modulation (QCIM). They also analyzed the BER performance of QCIM and compare it with other LoRa-based schemes.

There are only a few works on LoRa modulation that consider signal transmission from a single ED to GWs equipped with \emph{multiple antennas} \cite{nguyen2021performance, xu2019discrete}. In particular, the authors in \cite{nguyen2021performance} investigated performance improvement of the conventional LoRa modulation when multiple antennas are used at the GWs. They derived a BER expression for non-coherent detection in Rayleigh fading channels and proposed an iterative semi-coherent detection technique, whose performance is shown to approach that of the coherent detection without spending extra resources to estimate the channels. 

The works in \cite{tesfay2020multiuser, beltramelli2020lora, luvisotto2018use, haxhibeqiri2017lora, haxhibeqiri2018low, reynders2018improving} investigated the scalability of LoRa modulation in the downlink transmission from a single-antenna GW to multiple EDs. In particular, the authors in \cite{tesfay2020multiuser} developed a power-domain non-orthogonal multiple access (PD-NOMA) approach in order to increase the number of EDs that can be served in the downlink transmission of a LoRa network. They demonstrated that by exploiting the spread spectrum property of LoRa modulation, the use of successive interference cancelation, which is common in NOMA, is not needed at EDs, and hence, maintaining their low complexity implementation. In \cite{beltramelli2020lora}, the authors evaluated performance of ALOHA, slotted-ALOHA, and non-persistent carrier-sense multiple access (CSMA) when used with LoRa modulation, and showed the tradeoff among various random multiple access techniques and parameters of LoRa modulation, such as spreading factor, and the number of EDs and coverage area.

Against the above literature, in this paper, we investigate a new model to improve the uplink scalability of LoRa networks by allowing concurrent transmission (in the same time slot) of multiple EDs to multiple multi-antenna GWs using the same spreading factor and over the same frequency band. In essence, such a system design embraces collision rather than suffering from it. To make it work, the most important task is to develop a detection algorithm that can jointly detect the information simultaneously (concurrently) transmitted by multiple EDs, an approach known as \emph{multiuser detection}. To this end, we first derive the maximum likelihood (ML) decision rule for the non-coherent multiuser detection. Since the ML detection rule turns out to be computationally prohibitive for practical LoRa networks, we then develop a two-stage sub-optimal detection algorithm to balance the computational complexity and detection performance. In the first stage, we identify the transmit chirps without knowing which EDs transmit the identified chirps. In the second stage, we determine the EDs that transmit the specific chirps that were identified in the first stage. To improve the detection performance of the second stage, we optimize the transmit power of EDs to reduce the similarity, measured by the Jaccard coefficient, between the received powers of any pair of EDs. We show that the power control optimization problem is non-convex, and hence, hard to solve. We use concepts from successive convex approximation to transform the non-convex problem to an approximate convex one that can be solved iteratively and guaranteed to reach a sub-optimal solution.
Simulations results demonstrate the merits of the proposed network model, the importance of power control, and the effectiveness of the two-stage sub-optimal detection algorithm.

The remainder of this paper is organized as follows. Section \ref{sec:model} introduces the system model. Section \ref{sec:ML} derives the non-coherent ML multiuser detection rule. The proposed two-stage sub-optimal detection algorithm is presented in Section \ref{sec:sol}. The power control optimization problem is formulated and solved in Section \ref{sec:power}. Simulation results are provided in Section \ref{sec:sim}. Section \ref{sec:con} concludes the paper.

\section{System Model}\label{sec:model}

We consider the uplink transmission of a LoRa network in which $N_u$ single-antenna EDs communicate by means of chirp spread spectrum (CSS), or LoRa modulation with $L$ gateways (GWs), each equipped with $N_t$ antennas. Different from a conventional LoRa network, here all the $N_u$ EDs are allowed to transmit simultaneously (i.e., in the same time slot) using the same frequency band and spreading factor. Thus, compared to the conventional LoRa network in which no more than one ED can transmit in the same time slot over the same frequency band and using the same SF, the considered network can theoretically accommodate $N_u$ times more devices in the same service area.

Let $W$ and $T_{\mathrm{sym}}$ denote,  respectively, the bandwidth and symbol duration of LoRa modulation. Then with the sampling period of $T_s= 1/W$, the number of samples in each LoRa symbol (i.e., chirp) is given as $M = T_{\mathrm{sym}}/T_s = 2^{\rm SF}$, where ${\rm SF} \in \{7, 8, \ldots, 12\}$ is the spreading factor, which is also the number of information bits that can be carried by one LoRa chirp. The basic baseband up chirp is made up of the following $M$ time samples \cite{hanif2020slope}:
%%%%%%%%%%%%
\begin{equation}
x_0[n]=\exp\left\{j2\pi\left(\frac{ n^2}{2M}-\frac{n}{2}\right)\right\},\; n=0,1\dots, M-1.
\end{equation}
%%%%%%%%%%%%%
From this basic chirp, a set of $M$ orthogonal chirps can be generated as:
%%%%%%%%%%%%%%%
\begin{equation}
x_m[n]=x_0[n+m],\; m=0,1\dots, M.
\end{equation}
%%%%%%%%%%%%%%%%

By allowing $N_u$ EDs to transmit simultaneously in the same frequency band and using the same SF, the $N_t \times 1$ received signal vector at the $\ell$th GW is given as:
%%%%%%%%%%%%%%
\begin{equation}\label{eq3}
\boldsymbol{y}_{\ell}[n]=\sum_{g=1}^{N_u}\boldsymbol{h}_{g,\ell}\sqrt{p_g}x_{m_g}[n]+\boldsymbol{w}_{\ell}[n],
\end{equation}
%%%%%%%%%%%%%%%
where $p_g$ denotes the transmit power of the $g$th ED, $g = 1, \ldots, N_u$, $x_{m_g}[n]$ is the chirp transmitted by the $g$th ED (i.e., the transmitted symbol is $m_g$, where $m_g \in \{0,1,\dots, M-1\}$), $\boldsymbol{w}_{\ell}[n]\sim\mathcal{CN}\left(0,\sigma^{2}\mathbf{I}_{N_t}\right)$ is the vector of AWGN samples, and $\boldsymbol{h}_{g,\ell}\sim\mathcal{CN}\left(0,\beta_{g,\ell}\mathbf{I}_{N_t}\right)$ is the vector of $N_t$ uncorrelated Rayleigh fading channel gains between the $g$th ED and the $\ell$th GW, and $\beta_{g,\ell}$ represents the large-scale fading. The channel is assumed to stay constant within a coherence time of $T_c\gg T_{\mathrm{sym}}$. %In \eqref{eq3}, the $g$th ED transmits the $m_g$th chirp, where $m_g \in \{0,1,\dots, M-1\}$.

Allowing $N_u$ EDs to share the same bandwidth and SF causes inter-device interference, which needs to be properly handled at the network server. Suppressing the inter-device interference can be effectively done with coherent detection and by leveraging the large number of antennas at the GWs in order to achieve the asymptotic orthogonality property of the wireless channels among different EDs. Unfortunately, coherent detection requires extra time/frequency and processing resources for explicit channel estimation. Moreover, in order to have good channel estimation quality, the pilot power needs to be relatively high as compared to the noise level, which can be challenging in LoRa networks since EDs are typically battery operated and are expected to last for several years. Therefore, the main objective of this paper is to develop a \emph{non-coherent} detection algorithm to recover the information bits transmitted simultaneously from multiple EDs. The proposed detection algorithm takes advantage of the large number of antennas at each gateway and are presented in detail in the next sections.

\section{Non-Coherent Maximum Likelihood Detection}\label{sec:ML}

With sufficient antenna spacing at each GW, it is reasonable to assume that the channels from an ED to the $N_t$ antennas of a given GW are independent while experiencing the same large-scale fading. Under such an assumption, it is convenient to drop the antenna index in the following analysis. Specifically, the received signal at an arbitrary antenna of the $\ell$th GW is given as:
%%%%%%%%%%
\begin{equation}
y_{\ell}[n]=\sum_{g=1}^{N_u}{h}_{g,\ell}\sqrt{p_g}x_{m_g}[n]+{w}_{\ell}[n].
\end{equation}
%%%%%%%%%%%%

The first step in detecting LoRa signals is to perform dechirping, i.e., multiplying the received signal at a given antenna with the conjugate of the basic chirp. Accordingly, the dechirped signal corresponding to an arbitrary antenna is given as \cite{ghanaatian2019lora}:
%%%%%%%%%%%%%%%%
\begin{align}
z_{\ell}[n] & = y_{\ell}[n]x_0^{*}[n] \nonumber \\
&=\Bigg[\sum_{g=1}^{N_u}{h}_{g,\ell}\sqrt{p_g}\exp\left\{j2\pi\left(\frac{ (n+m_g)^2}{2M}-\frac{n+m_g}{2}\right)\right\} \nonumber \\ & \hspace{2.5cm} +{w}_{\ell}[n]\Bigg] \times\exp\left\{-j2\pi\left(\frac{ n^2}{2M}-\frac{n}{2}\right)\right\}, \nonumber\\
&=\sum_{g=1}^{N_u}\underbrace{{h}_{g,\ell}\sqrt{p_g}\exp\left\{j2\pi\left(\frac{ m_g^2}{2M}-\frac{m_g}{2}\right)\right\}}_{\text{constant gain}} \nonumber \\ & \hspace{4.5cm}\underbrace{\exp\left\{\frac{j2\pi m_gn}{M}\right\}}_{\text{linear phase}}+\hat{w}_{\ell}[n],\nonumber\\
&=\sum_{g=1}^{N_u}\hat{h}_{g,\ell}\sqrt{p_g}\exp\left\{\frac{j2\pi m_gn}{M}\right\}+\hat{w}_{\ell}[n].\label{eq5}
\end{align}
%%%%%%%%%%%%%%%%%%
In \eqref{eq5}, we define $\hat{h}_{g,\ell} = {h}_{g,\ell}\exp\left\{j2\pi\left(\frac{ m_g^2}{2M}-\frac{m_g}{2}\right)\right\}$ and  $\hat{w}_{\ell}[n]  ={w}_{\ell}[n]x_0^{*}[n]\sim\mathcal{CN}(0,\sigma^2)$. Next, we apply the $M$-point discrete Fourier transform (DFT) on the set of $M$ dechirped signal samples $z_{\ell}[n]$, $n=0,\ldots,M-1$, and obtain
%%%%%%%%%%%%%%%%%%%5
\begin{IEEEeqnarray}{rCl}\label{eq6}
&&Z_{\ell}[k]=\frac{1}{\sqrt{M}}\sum_{n=0}^{M-1}z_{\ell}[n]\exp\left\{\frac{-j2\pi nk}{M}\right\}\nonumber\\
&&=\frac{1}{\sqrt{M}}\sum_{g=1}^{N_u}\hat{h}_{g,\ell}\sqrt{p_g}\sum_{n=0}^{M-1}\exp\left\{\frac{j2\pi n(m_g-k)}{M}\right\}+W_{\ell}[k]\nonumber \\
&&=\sqrt{M}\sum_{g\in\mathcal{U}_k}\hat{h}_{g,\ell}\sqrt{p_g}+W_{\ell}[k], \quad k=0,1,\dots, M-1, \nonumber\\
\end{IEEEeqnarray}
%%%%%%%%%%%%%%%%%%%%%
where $\mathcal{U}_k=\left\{g\vert m_g=k\right\}$ is the set of all EDs that transmit the $k$th chirp (corresponding to the $k$th frequency bin after the DFT). The last identity in \eqref{eq6} follows from the fact that the second sum in the line above equals to $M$ when $m_g=k$ and zero when $m_g \neq k$.

Equation \eqref{eq6} gives the values across the $M$ frequency bins corresponding to an arbitrary single antenna at the $\ell$th GW. Collecting these observations over the entire antenna array of the $\ell$th GW yields an $N_t\times 1$ vector for the $k$th frequency bin as:
%%%%%%%%%%%%%%%%%%%%%%
\begin{equation}\label{eq7}
\boldsymbol{Z}_{\ell}[k]=\sqrt{M}\sum_{g\in\mathcal{U}_k}\hat{\boldsymbol{h}}_{g,\ell}\sqrt{p_g}+\boldsymbol{W}_{\ell}[k], \; k=0,1,\dots, M-1,
\end{equation}
%%%%%%%%%%%%%%%%%%%%%%%
where $\hat{\boldsymbol{h}}_{g,\ell}\sim\mathcal{CN}\left(0,\beta_{g,\ell}\mathbf{I}_{N_t}\right)$ since it is simply a phase-rotated version of ${\boldsymbol{h}}_{g,\ell}$. As a result,
$\boldsymbol{Z}_{\ell}[k]$ is a vector of random variables with distribution $\boldsymbol{Z}_{\ell}[k]\sim\mathcal{CN}\left(0,\left(M\sum_{g\in\mathcal{U}_k}p_g\beta_{g,\ell}+\sigma^2\right)\mathbf{I}_{N_t}\right)$.

Since non-coherent detection of LoRa signals is based on the received signal powers at different frequency bins, we define $r_{\ell,k}=\boldsymbol{Z}_{\ell}^H[k]\boldsymbol{Z}_{\ell}[k]$ as the total received power at the $\ell$th GW over the $k$th frequency bin, where $(\cdot)^H$ refers to the Hermitian operator. Collecting the signal powers in all $M$ frequency bins, we define $\boldsymbol{r}_{\ell}=[r_{\ell,0},r_{\ell,1},\dots, r_{\ell,M-1}]^{\top}$. Also for convenience, we represent all the symbols transmitted by the $N_u$ EDs as $\boldsymbol{m}=\left[{m_1},{m_2},\dots, {m_{N_u}}\right]$.

The development of the ML detection algorithm in this section exploits the large number of antennas at each GW. Specifically, as  $N_t$ tends to infinity, $r_{\ell,k}$ converges to a sum of average signal powers from the EDs transmitting the $k$th chirp, plus the noise power, i.e.,
%%%%%%%%%%%%%%%%%%%%%%%%%%%%%%
\begin{equation}\label{eq8}
\frac{1}{N_t}r_{\ell,k}\xrightarrow[N_t\rightarrow\infty]{\mathrm{a.s}}M\sum_{g\in\mathcal{U}_k}\beta_{g,\ell}p_g+\sigma^2\triangleq\rho_{\ell,k}.
\end{equation}
%%%%%%%%%%%%%%%%%%%%%%%%%%%%%
In fact, one can show that $r_{\ell,k}=\boldsymbol{Z}_{\ell}^H[k]\boldsymbol{Z}_{\ell}[k]$ is a Chi-square random variable with $2 N_t$ degrees of freedom with the following probability density function
%%%%%%%%%%%%%%%%%%%%%%%%%%%
\begin{equation}\label{eq9}
f(r_{\ell,k})=\frac{1}{(\rho_{\ell,k})^{N_t}(N_t-1)!}r_{\ell,k}^{N_t-1}\exp\left\{-\frac{r_{\ell,k}}{\rho_{\ell,k}}\right\}.
\end{equation}
%%%%%%%%%%%%%%%%%%%%%%%%%%

Using the fact that the power vectors $\{\boldsymbol{r}_{\ell}\}_{\ell=1}^L$ are statistically independent across the $L$ gateways, the likelihood function of power vectors $\{\boldsymbol{r}_{\ell}\}_{\ell=1}^L$ conditioned on the transmit symbols $\boldsymbol{m} = [m_1, m_2, \ldots, m_{N_u}]$ can be written as
%%%%%%%%%%%%%%%%%%%%%%%%%
\begin{equation}
\begin{split}
&\mathfrak{L}(\boldsymbol{r}_1,\boldsymbol{r}_2,\dots,\boldsymbol{r}_L|\boldsymbol{m})
=\prod_{\ell=1}^{L}\prod_{k=0}^{M-1}f(r_{\ell,k})\\
&=\prod_{\ell=1}^{L}\prod_{k=0}^{M-1}\frac{1}{(\rho_{\ell,k})^{N_t}(N_t-1)!}r_{\ell,k}^{N_t-1}\exp\left\{-\frac{r_{\ell,k}}{\rho_{\ell,k}}\right\},
\end{split}
\end{equation}
%%%%%%%%%%%%%%%%%%%%%%%%%%
and the corresponding log-likelihood function is
%%%%%%%%%%%%
\begin{equation}
\begin{split}
&\hat{\mathfrak{L}}(\boldsymbol{r}_1,\boldsymbol{r}_2,\dots,\boldsymbol{r}_L|\boldsymbol{m})
=\mathrm{ln}\left(\mathfrak{L}(\boldsymbol{r}_1,\boldsymbol{r}_2,\dots,\boldsymbol{r}_L|\boldsymbol{m})\right)\\
&=\frac{1}{(N_t - 1)!}+\sum_{\ell=1}^{L}\sum_{k=0}^{M-1}\left(\mathrm{ln}\left(\frac{r_{\ell,k}^{N_t-1}}{\rho_{\ell,k}^{N_t}}\right)-\frac{r_{\ell,k}}{\rho_{\ell,k}}\right).
\end{split}
\end{equation}
%%%%%%%%%%%%%
It is pointed out that the dependence of the likelihood function (and log-likelihood function) on $\boldsymbol{m}$ is through the sets $\{\mathcal{U}_k\}_{k=1}^M$, which determine the values of $\rho_{\ell,k}$ as defined in \eqref{eq8}.

Since $\ln(\cdot)$ is a monotonically increasing function and $\frac{1}{(N_t - 1)!}$ is a constant that is independent of the transmit symbols, the non-coherent ML detection of the transmitted $N_u$-tuple can be expressed as:
%%%%%%%%%%%%%%%%%%%%%
\begin{equation}\label{eq12}
\begin{split}
\hat{\boldsymbol{m}}&=\underset{\boldsymbol{m}\in \mathcal{S}_0}{\text{arg max}}\quad\hat{\mathfrak{L}}(\boldsymbol{r}_1,\boldsymbol{r}_2,\dots,\boldsymbol{r}_L|\boldsymbol{m})\\
&=\underset{\boldsymbol{m}\in\mathcal{S}_0}{\text{arg max}}\quad\sum_{\ell=1}^{L}\sum_{k=0}^{M-1}\left(\mathrm{ln}\left(\frac{r_{\ell,k}^{N_t-1}}{\rho_{\ell,k}^{N_t}}\right)-\frac{r_{\ell,k}}{\rho_{\ell,k}}\right),
\end{split}
\end{equation}
%%%%%%%%%%%%%%%%%%%%%
where $\mathcal{S}_0$ denotes the set of all possible $N_u$-tuples.

While finding the optimal solution for the ML detection problem in \eqref{eq12} is conceptually simple, the exhaustive search over the set $\mathcal{S}_0$ requires a complexity in the order of $M^{N_u}$, which is clearly prohibitive for $M = 2^{\rm SF}$ with ${\rm SF} = \{7, 8, \ldots, 12\}$ used in practical LoRa networks. Hence, in the next section we shall present a sub-optimal low-complexity detection algorithm. The key idea of our proposed detection algorithm is to reduce the search space by making use of the observation that since we have at most $N_u$ EDs that can transmit at any given time, there will be \textit{at most} $N_u \ll M$ chirps transmitted in every symbol duration.

\section{Proposed Low-Complexity Detection Algorithm}\label{sec:sol}

As discussed earlier, the search space can be reduced by noting that for the simultaneous uplink transmission of $N_u$ EDs  where each ED can transmit one chirp out of $M$ possible chirps, we will have at most $N_u \ll M$ chirps transmitted every symbol duration. However, it is possible that the same chirp is transmitted by more than one ED. Thus, the number of different transmitted chirps, denoted by $i$, that are transmitted by $N_u$ EDs satisfies $1\leq i\leq N_u$. The proposed sub-optimal detection algorithm solves the multiuser detection problem in \eqref{eq12} by decoupling it into two sub-problems that are solved in two stages. In the first stage, the algorithm identifies the $i$ transmitted chirps ($1\leq i \leq N_u \ll M$) out of the $M$ possible chirps without knowing which EDs transmit which identified chirps. In the second stage, the algorithm determines which EDs that transmit which chirps out of $i$ chirps that are identified from the first stage. Clearly, the size of the search space in the second stage is $i^{N_u}$, which is much smaller than the size $M^{N_u}$ of $\mathcal{S}_0$.

\subsection{First Stage: Identifying the Transmitted Chirps}

With non-coherent detection, it is expected that identifying the transmitted chirps amounts to identify the active frequency bins based on the powers calculated at all the gateways. As shown in \eqref{eq8}, as the number of antennas $N_t$ at each GW tends to infinity, the power $r_{\ell,k}$ provided by the antenna array of the $\ell$th GW over the $k$th frequency bin converges to $\rho_{\ell,k}$. It follows that the sum of $r_{\ell,k}$ over all the $L$ GWs at the $k$th frequency bin approaches to
%%%%%%%%%%%%%%%%%%%%%%%
%%%%%%%%%%%%%%%%%%%%%%%%%%%%%%
\begin{equation}\label{eq14}
\begin{split}
&\Upsilon[k] = {\frac{1}{N_t}}\sum_{\ell=1}^{L}r_{\ell,k} \xrightarrow[N_t\rightarrow\infty]{\mathrm{a.s}}
\sum_{\ell=1}^{L}\rho_{\ell,k}\\
&=\begin{cases}
M\sum_{\ell=1}^{L}\sum_{g\in\mathcal{U}_k}\beta_{g,\ell}p_g+L\sigma^2,\:\:\mathcal{U}_k\neq\emptyset,\\
L\sigma^2, \quad\text{otherwise},
\end{cases}
\end{split}
\end{equation}
%%%%%%%%%%%%%%%%%%%%%%%%%%%%%
where $\mathcal{U}_k=\left\{g\vert m_g=k\right\}$ is the set of all EDs that transmit the $k$th chirp.

The expression in \eqref{eq14} suggests that one can choose a power threshold of  $L\sigma^2$ to detect whether the $k$th chirp was transmitted, i.e., whether the $k$th frequency bin is \emph{active}. However, using such a threshold yields error-free detection only under the theoretical assumption of having an infinite number of antennas $N_t$ at each GW. Under a practical situation of having a limited (although can be very large) number of antennas at each gateway, a proper power threshold needs to be found. Specifically, the $k$th frequency bin is identified to be active or not by comparing $\Upsilon[k]$ with a certain power threshold  $P_{\mathrm{th}}$ as follows:
%%%%%%%%%%%%%%%%%%%%
\begin{equation}\label{eqthr}
\Upsilon[k] \overset{\text{active}}{\underset{\text{inactive}}{\gtrless}} P_{\mathrm{th}}, \quad k = 0, 1, \ldots, M-1.
\end{equation}
%%%%%%%%%%%%%%%%%%%%%
Finding the value of $P_{\mathrm{th}}$ is crucial to the successful identifications of the active frequency bins, and this is what we discuss in the rest of this subsection.

We showed in \eqref{eq9} that $r_{\ell,k}$ is a Chi-square distributed random variable with $2N_{t}$ degrees of freedom. For a large number of antennas at the GWs, $r_{\ell,k}/N_t$  can be approximated as a  Gaussian random variable with mean $\rho_{\ell,k}$ and variance $\rho_{\ell,k}^2/N_t$. Then it follows that
%%%%%%%%%%%%%%%%%%
\begin{equation}\label{eq:dist_Upsilon}
\Upsilon[k]\sim\mathcal{CN}\left(\sum_{\ell=1}^{L}\rho_{\ell,k},\sum_{\ell=1}^{L}\rho_{\ell,k}^2/N_t\right).
\end{equation}
%%%%%%%%%%%%%%%%%%%%%
Equation \eqref{eq:dist_Upsilon} reveals that  the mean and variance of $\Upsilon[k]$ depend on whether the $k$th frequency bin is active or not. Furthermore, in case the $k$th frequency bin is active, it depends on the number of EDs that transmit the $k$th chirp.

With concurrent transmission of $N_u$ EDs, we know that there are at most $N_u \ll M$ transmitted chirps in every symbol duration, and hence, there are at most $N_u$ active frequency bins. Therefore, for a given $P_{\mathrm{th}}$, the decision rule to identify the active frequency bins is as follows:
 \begin{itemize}
  	\item  If there are $i\leq N_u$ frequency bins having powers higher than $P_{\mathrm{th}}$, all of the $i$ frequency bins are identified as active.
  	\item If there are $i>N_u$ frequency bins having powers higher than $P_{\mathrm{th}}$, only the $N_u$ frequency bins with the highest powers are identified as active.
\end{itemize}
Using the above decision rule, we can find $P_{\mathrm{th}}$ to minimize the error probability (i.e., the false alarm probability), which is equivalent to maximizing the probability of detecting the active bins correctly.

Let $N^{+}$, $1\leq N^{+}\leq N_u$, denote the number of different chirps sent by $N_u$ EDs. Then, the events of correct identification of $N^{+}$ are as follows:
%%%%%%%%%%%%%%%
\begin{itemize}
\item $N^{+}=N_u$ chirps are transmitted, and $\Upsilon[k]$ on all the $N_u$ true active frequency bins are above $P_{\mathrm{th}}$ and also greater than the powers on all other $M - N_u$ true inactive frequency bins (note that it is possible to have the measured power on a true inactive bin higher than $P_{\mathrm{th}}$).
\item $N^{+}=i<N_u$ chirps are transmitted, and $\Upsilon[k]$ on all the $i$ true active frequency bins are above $P_{\mathrm{th}}$, whereas the powers on all the $M - i$ inactive frequency bins are below $P_{\mathrm{th}}$.
\end{itemize}
Then, the probability of correct identification of the $N^{+}$ active frequency bins can be calculated as
%%%%%%%%%%%%%%%%%%%
\begin{equation}\label{eq17}
\begin{split}
P_{\mathrm{correct}}&=P({\rm correct}|N^{+}=N_u)P(N^{+}=N_u)\\
&+\sum_{i=1}^{N_u-1}P({\rm correct}|N^{+}=i)P(N^{+}=i).
\end{split}
\end{equation}
%%%%%%%%%%%%%%%%%%%
The computation of each term in the above expression is as follows.

First, the case $N^{+}=N_u$ means that the chirps sent by $N_u$ EDs are all different. In other words, the set  {$\mathcal{U}_k$} corresponding to each active frequency bin includes one ED only. Without loss of generality, we index the active frequency bins from $g=1$ to $g=N_u$ and use the random variable $\Upsilon_g$ to denote the power calculated in each of these active frequency bins. It is noted that, while the actual powers calculated over active frequency bins are likely different in each symbol duration, all the active bins' powers have the same statistics and can be characterized  by the random variable $\Upsilon_g$. Specifically, it follows from {\eqref{eq14}} that {$\Upsilon_g \sim f_{\Upsilon_g}=\mathcal{CN}(L\mu_g,L\sigma_g^2/N_t)$}, where
\begin{equation}\label{eq23}
\mu_{g}= \frac{1}{L}\sum_{\ell=1}^{L}\left(M\beta_{g,\ell}p_g+\sigma^2\right),
\end{equation}
%%%%%%%%%%%%%%%%%
and
%%%%%%%%%%%%%%%%%%%%%%
\begin{equation}\label{eq24}
\sigma_{g}^2=\frac{1}{L}\sum_{\ell=1}^{L}\left(M\beta_{g,\ell}p_g+\sigma^2\right)^2.
\end{equation}
Likewise, we use the random variable $\Upsilon_{\bar g}$ to denote the power calculated in each of the remaining $M-N_u$ inactive frequency bins. Then it follows from {\eqref{eq14}} that $\Upsilon_{\bar g}$ is a Chi-square distributed random variable with $2N_{t}L$ degrees of freedom.

Then, the probability $P({\rm correct}|N^{+}=N_u)$ can be calculated as:
%%%%%%%%%%%%%%%%%%%%%%%5
\begin{equation}\label{P-Nu}
\begin{split}
&P({\rm correct}|N^{+}=N_u)\\
&=\sum_{g=1}^{N_u}\int_{P_{\mathrm{th}}}^{\infty}\left[P\left(\text{inactive bin's power}<U_0\right)\right]^{M-N_u}\\
&\times f_{\Upsilon_g}\left(g\text{th active bin's power}=U_0\right)\\
&\times\prod_{q=1,q\neq g}^{N_u}P\left(q\text{th active bin's power}>U_0\right)\dd U_0.
\end{split}
\end{equation}
%%%%%%%%%%%%%%%%%%%%%%
The probability that the $g$th active frequency bin's power is higher than $U_0$ is
%%%%%%%%%%%%%%%%%%%%%%%
%%%%%%%%%%%%%%%
\begin{equation}\label{eq25}
\begin{split}
&P\left(g\text{th active bin's power}>U_0\right)=P(\Upsilon_g>U_0)\\
&=\frac{1}{\sqrt{2\pi L\sigma_{g}^2/N_t}}\int_{U_0}^{\infty}\exp\left\{-\frac{(U_0-L\mu_{g})^2}{2L\sigma_{g}^2/N_t}\right\}\dd U_0,\\
&=\frac{1}{2}-\frac{1}{2}\mathrm{erf}\left(\frac{U_0-L\mu_{g}}{\sqrt{2L\sigma_{g}^2/N_t}}\right).
\end{split}
\end{equation}
%%%%%%%%%%%%%%%%%
On the other hand, the probability that the powers in $M-N_u$ inactive frequency bins are all below an arbitrary value $U_0$ is
%%%%%%%%%%%%%%%
\begin{equation}\label{eq25b}
	\begin{split}
	&\left[P\left(\text{inactive bin's power}<U_0\right)\right]^{M-N_u}\\
&=\left[P\left(\Upsilon_{\bar g}<U_0\right)\right]^{M-N_u}\\	 &=\left(1-\exp\left(-\frac{U_0}{\sigma^2}\right)\sum_{q=0}^{N_tL-1}\frac{1}{q!}\left(\frac{U_0}{\sigma^2}\right)^q\right)^{M-N_u}.
	\end{split}
	\end{equation}
%%%%%%%%%%%%%%%%%
Substituting \eqref{eq23}, \eqref{eq24}, \eqref{eq25}, and \eqref{eq25b} into \eqref{P-Nu} yields
	%%%%%%%%%%%%%%%%%%
	\begin{equation}\label{eq30}
	\begin{split}
	&P({\rm correct}|N^{+}=N_u)\\
	&=\sum_{g=1}^{N_u}\frac{1}{\sqrt{2\pi L\sigma_{g}^2/N_t}}\int_{P_{\mathrm{th}}}^{\infty}\exp\left\{-\frac{(U_0-L\mu_{g})^2}{2L\sigma_{g}^2/N_t}\right\}\\
	 &\times\left(1-\exp\left(-\frac{U_0}{\sigma^2}\right)\sum_{q=0}^{N_tL-1}\frac{1}{q!}\left(\frac{U_0}{\sigma^2}\right)^q\right)^{M-N_u}\\
	&\times\prod_{q=1,q\neq g}^{N_u}\left[\frac{1}{2}-\frac{1}{2}\mathrm{erf}\left(\frac{U_0-L\mu_{g}}{\sqrt{2L\sigma_{{q}}^2/N_t}}\right)\right]\dd U_0.\\
	\end{split}
	\end{equation}
	%%%%%%%%%%%%%%%%%%%%

Next, for the case $N^{+}=i<N_u$, we have
%%%%%%%%%%%%%%%%%%%%%%%%%%
\begin{equation}\label{eq26}
\begin{split}
&P({\rm correct}|N^{+}=i)\\
&=\left[P\left(\text{inactive bin's power}<P_{\mathrm{th}}\right)\right]^{M-i}\\
&\times\prod_{q=1}^{i}P\left(q\text{th active bin's power}>P_{\mathrm{th}}\right).
\end{split}
\end{equation}
%%%%%%%%%%%%%%%%%%%%%%%%%

Unlike the case $N^{+}=N_u$ where each active frequency bin corresponds to only one ED and the mean and variance of the power variable $\Upsilon_g$ for each active frequency bin can be determined easily, the mean and variance of $\Upsilon[k]$ for each active frequency bin in the case $N^{+}=i<N_u$ depend on how the $N_u$ EDs share these $i$ active bins. Given the high complexity involved in obtaining an exact expression for $P({\rm correct}|N^{+}=i)$, we seek a simpler lower bound. To this  end, we identify the $i$ smallest values of $\mu_{g}$ in \eqref{eq23} (or equivalently $\sigma_{g}^2$ in \eqref{eq24}) over $g=1,\ldots,N_u$ and use them as the means and variances of power variables $\Upsilon[k]$ for the $i$ active frequency bins. Doing so leads to the following lower bound of \eqref{eq26}:
%%%%%%%%%%%%%%%%%%
\begin{equation}
	\begin{split}
	&P({\rm correct}|N^{+}=i)\geq P^{\rm(LB)}({\rm correct}|N^{+}=i)\\
 &=\left(1-\exp\left(-\frac{P_{\mathrm{th}}}{\sigma^2}\right)\sum_{q=0}^{N_tL-1}\frac{1}{q!}\left(\frac{P_{\mathrm{th}}}{\sigma^2}\right)^q\right)^{M-i}\\
 &\times\prod_{g\in\mathcal{S}_i}\left[\frac{1}{2}-\frac{1}{2}\mathrm{erf}\left(\frac{P_{\mathrm{th}}-L\mu_{g}}{\sqrt{2L\sigma_{g}^2/N_t}}\right)\right],\\
	\end{split}
	\end{equation}
	%%%%%%%%%%%%%%%%%%%
where $\mathcal{S}_i$ denotes the set of $i$ EDs giving the $i$ lowest values of $\mu_{g}$.

The only thing left to find the lower bound on the probability of correct identification in \eqref{eq17} is to calculate the probability of having $i$ chirps transmitted by $N_u$ EDs. It is given as
%%%%%%%%%%%%%%%%%%%%5
\begin{equation}
P(N^{+}=i)=\frac{C_i\binom{M}{i}}{M^{N_u}},\; i=1,\ldots, N_u,
\end{equation}
%%%%%%%%%%%%%%%%%%%%
where $C_i$ is the total number of possible $N_u$-tuples given $i$ chirps are transmitted. This quantity can be easily found in a recursive manner as outlined in Lemma 1 below.

\textbf{Lemma 1:} Given the identities of $i$ transmitted chirps, the total number of possible $N_u$-tuples can be calculated as
%%%%%%%%%%
\begin{subequations}\label{eq16}
	\begin{eqnarray}
	&&C_1=1,\\
	&&C_i=i^{N_u}-\sum_{k=1}^{i-1}C_k\binom{i}{k}.
	\end{eqnarray}
\end{subequations}
%%%%%%%%%%

\textit{Proof:} If all $N_u$ EDs transmit the same chirp, then the number of possible $N_u$-tuples is obviously $C_1=1$. If there are 2 chirps transmitted, an ED transmits either one of these 2 chirps. As a result, the number of possible $N_u$-tuples is $2^{N_u}$. By excluding the case that all EDs transmit the same chirp (either one of the two transmitted chirps), we have $C_2=2^{N_u}-1\times 2$.
Similarly, when 3 chirps are transmitted, the number of candidate $N_u$-tuples is equal $3^{N_u}$. By excluding the cases when all EDs transmit 1 or 2 chirps, we have $C_3=3^{N_u}-C_2\times\binom{3}{2}-C_1\times\binom{3}{1}$. Proceeding in the same way, one can   show that the number of possible $N_u$-tuples when $i$ chirps are transmitted can be calculated as in Lemma 1. \hfill$\blacksquare$

In summary, the probability of correct identification of the $N^{+}$ active frequency bins (which is equivalent to identifying the $N^{+}$ transmitted chirps) in \eqref{eq17} can be lower bounded as:
%%%%%%%%%%%%%%%
\begin{equation}
\begin{split}
P_{\rm correct}^{\rm(LB)}&=P({\rm correct}|N^{+}=N_u)P(N^{+}=N_u)\\
&+\sum_{i=1}^{N_u-1} P^{\rm(LB)}({\rm correct}|N^{+}=i)P(N^{+}=i).
\end{split}
\end{equation}
%%%%%%%%%%%%%%%

\begin{figure}[t!]
	\centering
	\mbox{\scalebox{0.325}{\includegraphics{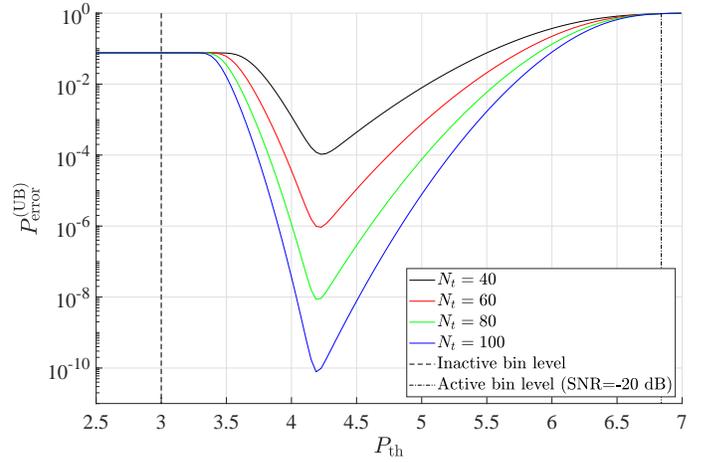}}}
	\caption{Error probability of detecting active bins (i.e., transmitted chirps), ${\rm SNR}=-20$ dB with 5 EDs.}\label{se}
\end{figure}

Obviously, the error probability in identifying the transmitted chirps is upper bounded by:
%%%%%%%%%%%%%%%%%%
\begin{equation}\label{eqP_UB}
P_{\rm error}^{\rm(UB)}=1-P_{\rm correct}^{\rm(LB)}.
\end{equation}
%%%%%%%%%%%%%%%%%%
In Fig. \ref{se}, we plot $P_{\rm error}^{\rm(UB)}$ versus $P_{\mathrm{th}}$ for different numbers of antennas at the GWs. As can be seen, $P_{\rm error}^{\rm(UB)}$ is a unimodal function of $P_{\mathrm{th}}$. Hence, the optimal value of $P_{\mathrm{th}}$ that minimizes the upper bound of the error probability in identifying the transmitted chirps can be easily found, for example by the Golden search.

\subsection{Second Stage: Detection of Chirps Transmitted by the EDs}

It can be seen from Lemma 1 that if the transmitted chirps have been identified, the total number of possible $N_u$-tuples is significantly smaller than that in the original problem in \eqref{eq12}. For example, if $N_u=5$, the number of possible $N_u$-tuples calculated according to Lemma 1 is $C_1=1$, $C_2=30$, $C_3=150$, $C_4=240$, and $C_5=120$, which are much smaller than $M^{5}$. Furthermore, the number of possible $N_u$-tuples in Lemma 1 is independent of $M$ and only depends on the number of EDs sharing the same frequency band and SF. It is clear that identifying the transmitted chirps plays a crucial role to reduce the complexity of the proposed detection algorithm.

After the active frequency bins have been identified from Stage 1, the original candidate set $\mathcal{S}_0$ in \eqref{eq12} is replaced with a reduced set, denoted by $\mathcal{S}_d$, which contains all possible $N_u$-tuples constructed from the identified (active) frequency bins from Stage 1. The ML multiuser detection problem in \eqref{eq12} then becomes:
%%%%%%%%%%%%%%%
%%%%%%%%%%%%%%%%%%%%%
\begin{equation}\label{eq27}
\begin{split}
\hat{\boldsymbol{m}}=\underset{\mathcal{S}_d}{\text{arg max}}\quad\sum_{\ell=1}^{L}\sum_{k=0}^{M-1}\left(\mathrm{ln}\left(\frac{r_{\ell,k}^{N_t-1}}{\rho_{\ell,k}^{N_t}}\right)-\frac{r_{\ell,k}}{\rho_{\ell,k}}\right).
\end{split}
\end{equation}
%%%%%%%%%%%%%%%%%%%%%
%%%%%%%%%%%%%

The tradeoff for reducing the complexity from searching over $\mathcal{S}_0$ to searching over $\mathcal{S}_d$ is performance degradation since a mistake in the active bin identification in Stage 1 may lead to error propagation and potentially remove the true transmitted $N_u$-tuple from the candidate set $\mathcal{S}_d$. However, from the simulation results, we will show that the performance gap between searching over the reduced set $\mathcal{S}_d$ and the original set $\mathcal{S}_0$ is small. Intuitively, when the active frequency bin identification in Stage 1 is wrong, it means there are inactive frequency bins having powers higher than the active frequency bins' powers. In such a case, even the detection based on the original candidate set $\mathcal{S}_0$ will likely be wrong.

Even with the reduced candidate set $\mathcal{S}_d$, the detection in \eqref{eq27} can be made a little simpler. Specifically, it can observed from \eqref{eq27} that the contributions of the {inactive} frequency bins to the log-likelihood function are identical for all transmitted $N_u$-tuples in the reduced set, and hence, can be removed. 
This observation leads to the following simpler detection rule:
%%%%%%%%%%%%%%%%%%%%%
\begin{equation}\label{eq28}
\begin{split}
\hat{\boldsymbol{m}}
&=\underset{\mathcal{S}_d}{\text{arg max}}\quad\sum_{\ell=1}^{L}\sum_{k\in\mathcal{S}_+}^{}\left(\mathrm{ln}\left(\frac{r_{\ell,k}^{N_t-1}}{\rho_{\ell,k}^{N_t}}\right)-\frac{r_{\ell,k}}{\rho_{\ell,k}}\right),
\end{split}
\end{equation}
%%%%%%%%%%%%%%%%%%%%%
where $\mathcal{S}_+$ is the set of active frequency bins  identified in Stage~1.
%%%%%%%%%%%%%%%%%

\section{Power Control}\label{sec:power}

Recall that the proposed sub-optimal detection algorithm consists of two stages: Stage 1 identifies active frequency bins without specifying associated EDs, whereas Stage 2 detect the EDs associated with the active frequency bins identified in Stage 1 by solving the simplified multiuser detection problem in \eqref{eq28}. It can be observed from \eqref{eq28} that the detection performance in Stage 2 strongly depends on the difference in the received power levels of the transmitting EDs on the identified active frequency bins at the $L$ GWs. More specifically, if the received powers of the EDs on the identified active frequency bins are dissimilar, it will be easier to detect these EDs. This observation suggests we can perform a suitable power control policy of the transmitting EDs to produce dissimilar received power levels at the $L$ GWs, and hence improving the performance of the multiuser detection in \eqref{eq28}.

\subsection{Power Control Problem Formulation}

To develop a suitable power control problem, we denote the expected bin power of the $g$th ED, $g=1,\ldots, N_u$, at the $L$ GWs as the $L\times 1$ vector $\boldsymbol{\hat{\mu}}_{g}=[\mu_{g,1},\mu_{g,2},\ldots,\mu_{g,L}]^{T}$, where $\hat{\mu}_{g,\ell}=M\beta_{g,\ell}p_g+\sigma^2$ if only the $g$th ED transmits on the identified active bin. The objective of the proposed power control policy is to minimize the similarity of the expected bin power vectors among different EDs, i.e., $\boldsymbol{\hat{\mu}}_g$ and $\boldsymbol{\hat{\mu}}_{g'}$, for $g \neq g'$. A possible choice to measure the similarity between the two vectors $\boldsymbol{\hat{\mu}}_g$ and $\boldsymbol{\hat{\mu}}_{g'}$ is the Jaccard coefficient \cite{cha2007comprehensive}, defined as:
\begin{equation}\label{eq:jaccard}
J_{g,g'}=\frac{\boldsymbol{\hat{\mu}}_g^T\boldsymbol{\hat{\mu}}_{g'}}{\left\Vert\boldsymbol{\hat{\mu}}_g\right\Vert^2+\left\Vert\boldsymbol{\hat{\mu}}_{g'}\right\Vert^2-\boldsymbol{\hat{\mu}}_g^T\boldsymbol{\hat{\mu}}_{g'}}, \quad g \neq g'.
\end{equation}
It can be verified that $J_{g,g'}$ lies between 0 to 1, with the value of 1 representing the highest similarity and the value of 0 representing the lowest similarity between the two vectors.

The Jaccard coefficient defined in \eqref{eq:jaccard} measures the similarity between the expected bin powers of two EDs $g$ and $g'$, $g \neq g'$, across the $L$ GWs. The definition of $\boldsymbol{\hat{\mu}}_g$ and $\boldsymbol{\hat{\mu}}_{g'}$ assumes that the two EDs  $g$ and $g'$ transmit two different chirps.
However, it is possible that two or more EDs transmit the same chirps, and if this scenario is not taken into account in the power control policy, errors may occur. This is because the contribution to the received powers at the GWs from one ED can be too weak compared to that of the other ED, and the detector may not realize that there are two transmitting EDs on the same frequency bin. The Jaccard coefficient when two EDs transmit the same chirp can be defined as
%%%%%%%%%%%%%%%%%%
\begin{equation}\label{eq:jaccard2}
J_{g,g'}^{\mathrm{(2)}}=\frac{\boldsymbol{\hat{\mu}}_{g,g'}^T\boldsymbol{\hat{\mu}}_{g'}}{\left\Vert\boldsymbol{\hat{\mu}}_{g,g'}\right\Vert^2+\left\Vert\boldsymbol{\hat{\mu}}_{g'}\right\Vert^2-\boldsymbol{\hat{\mu}}_{g,g'}^T\boldsymbol{\hat{\mu}}_{g'}},
\end{equation}
%%%%%%%%%%%%%%%%%%%%%%%%%
where $\boldsymbol{\hat{\mu}}_{g,g'}$ is the expected bin power of when both the $g$th and $g'$th EDs transmit the same chirp at the $L$ GWs. It is defined as $\boldsymbol{\hat{\mu}}_{g,g'}=[\mu_{g,g',1},\mu_{g,g',2},\ldots,\mu_{g,g',L}]^{T}$,  $g\neq g'$, where $\hat{\mu}_{g,g',\ell}=M\beta_{g,\ell}p_g+M\beta_{g',\ell}p_g'+\sigma^2$.

Taking both $J_{g,g'}$ and $J_{g,g'}^{\mathrm{(2)}}$ into account, the power control problem to improve the multiuser detection performance in \eqref{eq28} is formally expressed as
%%%%%%%%%%%%%%%%%%%%%%%%%
\begin{IEEEeqnarray}{rCl}
\mathcal{OP}_1: \underset{p_{g}}{\rm min} \quad \underset{g,g'}{\mathrm{max}} \quad & & \left\{J_{g,g'},   J_{g,g'}^{\mathrm{(2)}} \right\} \nonumber\\
\text{subject to} \quad &&
0\leq p_{g^{}}\leq p_{\rm max}, \: \forall g, \label{eq:power_budget}\\
%& J_{g,g'}^{-1}\geq \lambda\quad\forall g<g'\\
&& \frac{1}{L\sigma^2} \sum_{\ell=1}^{L}M\beta_{g,\ell}p_g \geq\epsilon,\: \forall g,\label{eq:snr}
\end{IEEEeqnarray}	
%%%%%%%%%%%%%%%%%%%%%%%%
where the constraint in \eqref{eq:power_budget} ensures that the ED's transmit power does not exceed a certain power budget $p_{\rm max}$, and the constraint in \eqref{eq:snr} guarantees that the average received SNR exceeds a predefined threshold $\epsilon$.

The optimization problem $\mathcal{OP}_1$ is non-convex because of the non-convexity of $J_{g,g'}$ and $J_{g,g'}^{\mathrm{(2)}}$. To facilitate obtaining a low-complexity sub-optimal solution, it is more convenient to work with its equivalent epigraph form. After a change of variables, $\mathcal{OP}_1$ can be reformulated as
%%%%%%%%%%%%%%%%%%%%%%%%
\begin{IEEEeqnarray}{rCl}\label{eqori}
\mathcal{OP}_2:	\underset{p_{g}, \lambda}{\rm max} \quad & & \lambda \nonumber\\
	\text{subject to} \quad && \eqref{eq:power_budget}, \eqref{eq:snr}, \nonumber\\
&& J_{g,g'}^{-1}\geq \lambda, \: \forall g<g', \label{eq:jac1}\\
&&  (J_{g,g'}^{\mathrm{(2)}})^{-1}\geq \lambda, \:\forall g<g'.\label{eq:jac2}
\end{IEEEeqnarray}		
%%%%%%%%%%%%%%%%%%%%%%%%
As can be seen in $\mathcal{OP}_2$, maximizing $\lambda$ in the objective function will maximize the lower bound of $J_{g,g'}^{-1}$, $g<g'$. This is equivalent to minimizing the upper bound of $J_{g,g'}$, $g<g'$, which enforces $J_{g,g'}$ to reduce its value. Similarly, maximizing $\lambda$ in the objective function also enforces $J_{g,g'}^{\mathrm{(2)}}$ to reduce its value. Hence, by solving $\mathcal{OP}_2$, we can reduce the similarities among the transmit power vectors of EDs, which eventually helps to improve the detection performance.

It is pointed out that, because of the non-convexity of the constraints in \eqref{eq:jac1} and \eqref{eq:jac2},  $\mathcal{OP}_2$ is still a non-convex optimization problem. In the following subsection, we propose a successive convex approximation technique to convert $\mathcal{OP}_2$ to a series of convex optimization problems, whose solutions are guaranteed to converge to a sub-optimal solution of $\mathcal{OP}_2$.

\subsection{Successive Convex Optimization}

The key step of the proposed successive convex optimization approach is to approximate the non-convex constraints in \eqref{eq:jac1} and \eqref{eq:jac2} with convex bounds at a feasible point. Then, we formulate an approximate convex optimization problem that can be solved efficiently in an iterative manner. The optimal solution of such an approximate convex optimization problem is guaranteed to be a feasible point of the original non-convex optimization problem \cite{Boyd2004convex}.

In the following, we explain the convex approximation of the constraint in \eqref{eq:jac1} in detail, and then  \eqref{eq:jac2} can be treated similarly. Let $\left\{\boldsymbol{p}_g^{(\kappa)},\bar{\lambda}^{(\kappa)}\right\}$ be the decision variables of the $\kappa$th iteration of the optimization problem in $\mathcal{OP}_2$. Then, we can rewrite \eqref{eq:jac1} as
%%%%%%%%%%%%%%%%%%%%%%
\begin{IEEEeqnarray}{rCl}\label{eq:1st_convex_const}
\frac{1}{\lambda^{(\kappa)}+3} \left\Vert\boldsymbol{\hat{\mu}_{g^{}}}+\boldsymbol{\hat{\mu}_{g'}}\right\Vert^2& \geq & \boldsymbol{\hat{\mu}}_g^T\boldsymbol{\hat{\mu}}_{g'}.
\end{IEEEeqnarray}
%%%%%%%%%%%%%%%%%%%%%%
The LHS of the above equation can be further simplified as
%%%%%%%%%%%%%%%%%%%%%%
\begin{equation}
\begin{split}
	 &\frac{\sum_{\ell=1}^{L}{\left(M\beta_{g,\ell}p_g^{(\kappa)}+M\beta_{g',\ell}p_{g'}^{(\kappa)}+2\sigma^2\right)^2}}{\lambda^{(\kappa)}+3} \\ & = \frac{\sum_{\ell=1}^{L}{\left(\theta_{g,g',\ell}^{(\kappa)}\right)}^2}{\lambda^{(\kappa)}+3},
\end{split}
\end{equation}
%%%%%%%%%%%%%%%%%%%%%%
where
\begin{equation}\label{eqC22}
\theta_{g,g',\ell}^{(\kappa)}=M\beta_{g,\ell}p_g^{(\kappa)}+M\beta_{g',\ell}p_{g'}^{(\kappa)}+2\sigma^2.
\end{equation}
Let $\left\{\boldsymbol{p}_g^{(\kappa - 1)},\bar{\lambda}^{(\kappa - 1)}\right\}$ be a feasible point to $\mathcal{OP}_2$ that can be found from solving the previous iteration of the optimization problem. Then, with the help of the following inequality
\begin{eqnarray}
\frac{x^2}{y} \geq \frac{2\bar{x}}{\bar{y}} x - \frac{\bar{x}^2}{\bar{y}^2}y, \quad \forall x, \bar{x}, y, \bar{y} >0,
\end{eqnarray}
we can show that
%%%%%%%%%%%%%%%%%%%%%%%
\begin{equation}\label{eq38}
\frac{{\left(\theta_{g,g',\ell}^{(\kappa)}\right)}^2}{\lambda^{(\kappa)}}\geq\frac{2{\theta}_{g,g',\ell}^{(\kappa - 1)}}{{\lambda}^{(\kappa - 1)}}\theta_{g,g',\ell}^{(\kappa)}-\frac{{\left(\theta_{g,g',\ell}^{(\kappa - 1)}\right)}^2}{
	{\left(\lambda^{(\kappa - 1)}\right)}^2}\lambda^{(\kappa)}.
\end{equation}
%%%%%%%%%%%%%%%%%%%%%%%
The RHS of \eqref{eq:1st_convex_const} can be rewritten as
\begin{IEEEeqnarray}{rCl}
&& p_g^{(\kappa)} p_{g'}^{(\kappa)}\sum_{\ell=1}^{L}M^2\beta_{g^{},\ell}\beta_{g',\ell} \nonumber\\ &&+\sum_{\ell=1}^{L}\sigma\left(M\beta_{g,\ell}p_g^{(\kappa)}+M\beta_{g',\ell}p_{g'}^{(\kappa)}+\sigma\right).\label{eq38a}
\end{IEEEeqnarray}
Using the inequality,
\begin{equation}
xy \leq \frac{\bar{x} \bar{y}}{4} \left(\frac{x}{\bar{x}} + \frac{y}{\bar{y}}\right)^2.
\end{equation}
a lower bound of \eqref{eq38a} is
\begin{IEEEeqnarray}{rCl}
&&\frac{{p}_g^{(\kappa - 1)} {p}_{g'}^{(\kappa - 1)}}{4}\left(\frac{p_g^{(\kappa)}}{{p}_g^{(\kappa-1)}}+\frac{p_{g'}^{(\kappa)}}{{p}_{g'}^{(\kappa-1)}}  \right)^2\sum_{\ell=1}^{L}M^2\beta_{g^{},\ell}\beta_{g',\ell}\nonumber\\
&&+\sum_{\ell=1}^{L}\sigma\left(M\beta_{g,\ell}p_g^{(\kappa)}+M\beta_{g',\ell}p_{g'}^{(\kappa)}+\sigma\right),\quad  0<g<g' \label{eq:RHS} \IEEEeqnarraynumspace
\end{IEEEeqnarray}

Finally, by substituting \eqref{eq38} and \eqref{eq:RHS} in \eqref{eq:1st_convex_const}, the non-convex constraint in \eqref{eq:jac1} is approximated as
%%%%%%%%%%%%%%%%%%%%%%%%%
\begin{IEEEeqnarray}{rCl}\label{eq40}
&&\sum_{\ell=1}^{L}\left(\frac{2{\theta}_{g,g',\ell}^{(\kappa - 1)}}{{\lambda}^{(\kappa - 1)}}\theta_{g,g',\ell}^{(\kappa)}-\frac{{\left(\theta_{g,g',\ell}^{(\kappa - 1)}\right)}^2}{{\left(\lambda^{(\kappa - 1)}\right)}^2}\lambda^{(\kappa)}\right) \nonumber\\
&&\geq \frac{{p}_g^{(\kappa - 1)} {p}_{g'}^{(\kappa - 1)}}{4}\left(\frac{p_g^{(\kappa)}}{{p}_g^{(\kappa-1)}}+\frac{p_{g'}^{(\kappa)}}{{p}_{g'}^{(\kappa-1)}}  \right)^2\sum_{\ell=1}^{L}M^2\beta_{g^{},\ell}\beta_{g',\ell}\nonumber\\
&&+\sum_{\ell=1}^{L}\sigma\left(M\beta_{g,\ell}p_g^{(\kappa)}+M\beta_{g',\ell}p_{g'}^{(\kappa)}+\sigma\right),\quad  0<g<g'. \IEEEeqnarraynumspace
\end{IEEEeqnarray}
%%%%%%%%%%%%%%%%%%%%%%%%%%
Similarly, \eqref{eq:jac2} can be approximated as
%%%%%%%%%%%%%%%%%%%%%%%%%
\begin{IEEEeqnarray}{rCl}\label{eq49}
	&&\sum_{\ell=1}^{L}\left(\frac{2{\theta}_{g,g',\ell}^{(2,\kappa - 1)}}{{\lambda}^{(\kappa - 1)}}\theta_{g,g',\ell}^{(2,\kappa)}-\frac{{\left(\theta_{g,g',\ell}^{(2,\kappa - 1)}\right)}^2}{{\left(\lambda^{(\kappa - 1)}\right)}^2}\lambda^{(\kappa)}\right) \nonumber\\
	&&\geq \frac{{p}_g^{(\kappa - 1)} {p}_{g'}^{(\kappa - 1)}}{4}\left(\frac{p_g^{(\kappa)}}{{p}_g^{(\kappa-1)}}+\frac{p_{g'}^{(\kappa)}}{{p}_{g'}^{(\kappa-1)}}  \right)^2\sum_{\ell=1}^{L}M^2\beta_{g^{},\ell}\beta_{g',\ell}\nonumber\\
	 &&+\sum_{\ell=1}^{L}\sigma\left(M\beta_{g,\ell}p_g^{(\kappa)}+M\beta_{g',\ell}p_{g'}^{(\kappa)}+\sigma\right),\quad  0<g<g',\IEEEeqnarraynumspace
\end{IEEEeqnarray}
%%%%%%%%%%%%%%%%%%%%%%%%%%
where
\begin{equation}\label{eqC22b}
\theta_{g,g',\ell}^{(2,\kappa)}= \left(M\beta_{g,\ell}p_g^{(\kappa)}+M\beta_{g',\ell}p_{g'}^{(\kappa)}+2\sigma^2\right).
\end{equation}

\textbf{Remark 1:} Given the large number of possible chirps, $M=2^{\rm SF}$, in a practical LoRa network, the probability of having $N_u$ EDs transmit $N_u$ different chirps is significantly higher than the probability of having less than $N_u$ chirps transmitted by $N_u$ EDs. As such, jointly optimizing $J_{g,g'}$ and $J_{g,g'}^{\mathrm{(2)}}$ with equal weights is not likely the best optimization strategy, and we should prioritize the optimization of $J_{g,g'}$ over the optimization of  $J_{g,g'}^{\mathrm{(2)}}$. This can be done by scaling $\theta_{g,g',\ell}^{(2,\kappa)}$ with a constant ${\alpha}$, $\alpha > 1$, i.e.,
\begin{equation}\label{eqC22c}
\theta_{g,g',\ell}^{(2,\kappa)}=\alpha \left(M\beta_{g,\ell}p_g^{(\kappa)}+M\beta_{g',\ell}p_{g'}^{(\kappa)}+2\sigma^2\right).
\end{equation}
Hence, starting with a feasible point $\left\{\boldsymbol{p}_g^{(\kappa - 1)},\bar{\lambda}^{(\kappa - 1)}\right\}$, the $\kappa$th instance of the optimization problem $\mathcal{OP}_2$ is modified to
\begin{IEEEeqnarray}{rCl}\label{eqop3}
	\mathcal{OP}_3^{(\kappa)}:	\underset{p_{g}^{(\kappa)}, \lambda^{(\kappa)}}{\rm max} \quad & & \lambda \nonumber\\
	\text{subject to} \quad && \eqref{eq:power_budget}, \eqref{eq:snr}, \eqref{eq40},  \eqref{eq49}, \eqref{eqC22}, \eqref{eqC22c}. \nonumber
\end{IEEEeqnarray}		
By solving the $\kappa$th instance of the convex optimization problem $\mathcal{OP}_3^{(\kappa)}$ using convex optimization tools, e.g., CVX \cite{grant2010cvx}, we find an optimal solution $\left\{\boldsymbol{p}_g^{(\kappa)},\bar{\lambda}^{(\kappa)}\right\}$ to $\mathcal{OP}_3^{(\kappa)}$ which is also a feasible point to the non-convex problem in $\mathcal{OP}_2$. The process repeats until convergence. The sequence of feasible points $\left\{\boldsymbol{p}_g^{(\kappa)},\bar{\lambda}^{(\kappa)}\right\}$ is guaranteed to improve the objective function of $\mathcal{OP}_2$ and will eventually converge to a local optimum point that satisfies the KKT conditions \cite{Boyd2004convex, nguyen2020max}.

The proposed non-coherent sub-optimal detection algorithm is summarized in \textbf{Algorithm 1}.
%%%%%%%%%%%%%%%%%%%%
\begin{algorithm}
	\caption{Proposed power control and two-stage multiuser detection}
	{\small
		\begin{algorithmic}[1]\label{al:al2}
			\REQUIRE Large scale fading coefficients $\beta_{g,\ell},p_{\mathrm{max}}$, and $\epsilon$.
			%\textit{Power control phase (offline)}		
			\STATE $\kappa=1$. \COMMENT Power control phase (offline)
			\STATE Initially set $p_g^{(\kappa)}=p_{\mathrm{max}}$, $\forall g$.
			\WHILE{Until convergence}
			\STATE Approximate constraints \eqref{eq40} and \eqref{eq49} at $p_g^{(\kappa)}$.
			\STATE Solve $\mathcal{OP}_3^{(\kappa)}$.
			\STATE Obtain new $p_g$, $\forall g$.
			\ENDWHILE
			\RETURN $p_{g}$.
			\STATE Calculate $P_{\mathrm{th}}$ with Golden search based on \eqref{eqP_UB}.\\
			%\textit{Two-stage detection phase}
			\STATE Identify set of active bins $\mathcal{S}_{+}$. \COMMENT Two-stage detection phase
			\STATE Build set $\mathcal{S}_d$ of all $N_u$-tuples based on $\mathcal{S}_{+}$.
			\STATE Calculating the likelihood of all candidates in $\mathcal{S}_d$ using \eqref{eq28}.
		    \STATE Choose the candidate with the maximum likelihood as $\hat{\boldsymbol{m}}$.
			\RETURN $\hat{\boldsymbol{m}}$.
		\end{algorithmic}
	}
\end{algorithm}
%%%%%%%%%%%%%%%%%%%%%%

\section{Simulation Results}\label{sec:sim}
\begin{figure}[thb!]
	\centering
	\includegraphics[width=0.45\textwidth]{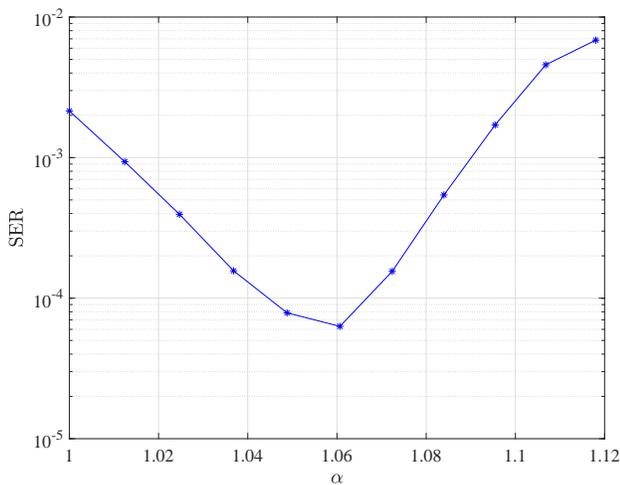}
	\caption{SER versus $\alpha$ with $N_u=3$ EDs, $N_t = 40$ antennas, and reference ${\rm SNR} = - 13$ dB.}\label{f5}
\end{figure}

In this section, we evaluate performance of the proposed multiuser detection algorithm in terms of the symbol error rate (SER). We consider a LoRa network with $L=3$ GWs simultaneously serving $N_u$ EDs that are randomly located inside a circle having a radius of 4 kilometers. The GWs are equally spaced on a circle of 2-kilometer radius from the center of the coverage area. To make the locations of EDs distinguishable, the minimum distance between any two EDs is set to be 500 meters. We consider Rayleigh fading channels with the large-scale fading coefficient calculated for the non-light-of-sight in-car model as in \cite{petajajarvi2015coverage}. Specifically, $\beta_{g,\ell} =  128.95 +23.2 \mathrm{log}_{10}d_{g,\ell}+z_{g,\ell}$ dB, where $d_{g,\ell}$ is the distance (in kilometers) from the $g$th ED to the $\ell$th GW, and $z_{g,\ell}$ represents shadow fading.  In all simulation scenarios, $\alpha$ is set to 1.061, except for Fig. \ref*{f5}, where we examine the impact of $\alpha$ on the system's performance. Other simulation parameters are summarized in Table \ref{tab:title}.

	\begin{table}[!t]
	\caption{Simulation parameters.} \label{tab:title}
	\begin{center}
		\begin{tabular}{|c|c|}
			\hline
			Parameter & Value \\
			\hline\hline
			GW height & 70 m \\
			\hline
			ED-GW minimum distance & 50 m \\
			\hline
			ED-ED minimum distance & 500 m \\
			\hline
			Bandwidth & 125 kHz \\
			\hline
			Spreading factor & 7  \\
            \hline
			Shadowing standard deviation & 7.8 dB \\
			\hline
			Noise figure & 6 dB \\
			\hline
		\end{tabular}
	\end{center}
\end{table}

In this section, the SER performance of the LoRa network with concurrent transmission of multiple EDs and under the proposed multiuser detection algorithm is investigated and compared to that of a conventional LoRa network in which a single ED communicates with a multiple-antenna GW \cite{nguyen2021performance}. In order to have a meaningful comparison between the two network models (single ED versus concurrent multiple EDs), a constraint is applied to the sum power of multiple EDs that are grouped for concurrent transmission. In detail, the transmit power $p_g^{\mathrm{(SU)}}$ of each ED in the proposed network model to achieve a predetermined SNR (called a reference SNR) at the closest GW is calculated. Then, the sum transmit power of the grouped EDs is constrained as:
%%%%%%%%%%%%%%%%%
\begin{equation}
\sum_{g=1}^{N_u}p_g\leq\sum_{g=1}^{N_u}p_g^{\mathrm{(SU)}}.
\end{equation}
%%%%%%%%%%%%%%%%%
On the other hand, from the reference SNR value, the SER in the case of single ED transmission can be theoretically calculated as in \cite{nguyen2021performance} when the optimal non-coherent detector is used. In this way the difference in the SER between the two network models can be observed at the same value of reference SNR.

\begin{figure}[t!]
	\centering
	\includegraphics[width=0.5\textwidth]{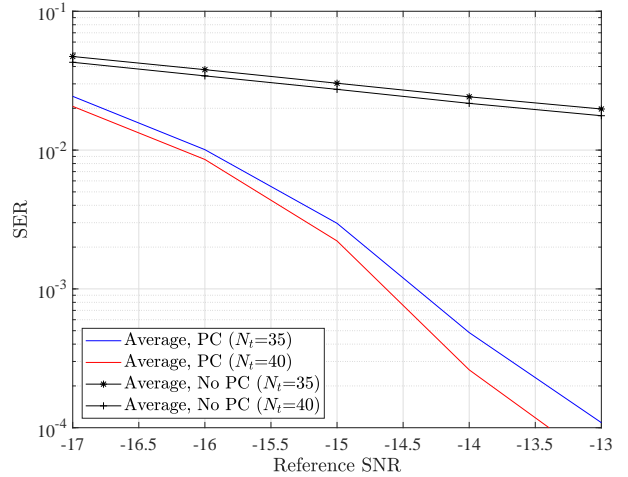}
	\caption{Effect of power control on SER for $N_u=3$ EDs, with $N_t = 35$ and $N_t = 40$.}\label{f3}
\end{figure}

First, to show the importance of power control in the proposed network model, Fig. \ref{f3} plots the SERs obtained with the proposed sub-optimal detection algorithm when $N_u=3$ EDs are grouped for concurrent transmission, and with and without power control. As can be seen, without power control, the SER is very high. As discussed before, the reason for this is that the differences among the EDs' expected power levels are small and the GWs cannot distinguish them despite increasing the number of antennas. On the contrary, with power control, a much better performance is achieved for the same total sum power. Because of the importance of power control, all the remaining results in this section are obtained with power control.
%%%%%%%%%%%%%%%%%%
%%%%%%%%%%%%%%%%

\begin{figure}[thb!]
	\centering
	\includegraphics[width=0.5\textwidth]{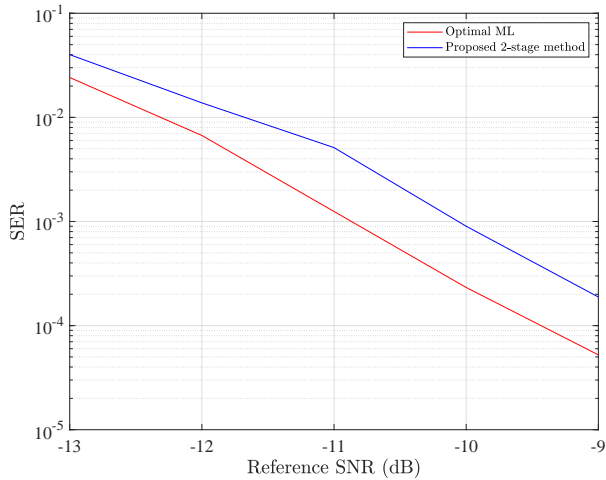}
	\caption{Performance comparison between the optimal ML detection and the proposed sub-optimal detection for $N_u = 2$ EDs operating at ${\rm SF} = 5$, and with $N_t = 35$.}\label{f_opt_vs_subopt}
\end{figure}

Fig. \ref{f_opt_vs_subopt} compares performance of the optimal ML multiuser detection (see \eqref{eq12}) and the proposed sub-optimal detection algorithm for $N_u=2$ EDs operating at $\rm SF$ = 5 and with $N_t = 35$. Note that these parameter values are chosen to enable the implementation of the ML detection. As pointed out before, with SF values of 7 to 12 in a practical LoRa network, the complexity of the ML detection is simply prohibitive. As can be seen from the figure, the proposed sub-optimal detection algorithm performs within 1 dB of the optimal ML detection. This clearly demonstrates the effectiveness of the proposed sub-optimal detection in balancing detection performance and computational complexity.
%%%%%%%%%%%%%%%%%%
%%%%%%%%%%%%%%%%

\begin{figure}[thb!]
	\centering
	\includegraphics[width=0.5\textwidth]{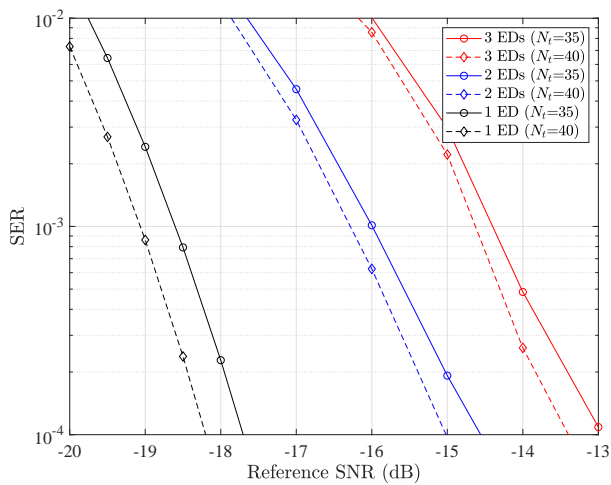}
	\caption{Average SER versus reference SNR for different numbers of EDs: $N_t = 35$ and $N_t = 40$.}\label{f1}
\end{figure}

Fig. \ref{f1} plots the average SER curves versus the reference SNR of the proposed network model for the cases of $N_u=2$ and $N_u=3$ and for two different sizes of the antenna array at each GW, namely $N_t = 35$ and $N_t = 40$. Also plotted in the figures are the SER curves of the conventional network model with single ED transmission. It can be seen that, for both cases of $N_t = 35$ and $N_t = 40$ antennas at each GW, in order to achieve a SER of $10^{-4}$, the proposed multiuser LoRa networks with 2 and 3 EDs transmitting concurrently require about 3.0 and 4.7 dB more in the transmit power, respectively. Given that the overall network capacity can be doubled or tripled by letting $N_u=2$ or $N_u=3$ EDs transmit concurrently and detecting their information jointly, such transmit power penalties can be well justified.

%%%%%%%%%%%%%%%%%%
%%%%%%%%%%%%%%%%

\begin{figure}[thb!]
	\centering
	\includegraphics[width=0.5\textwidth]{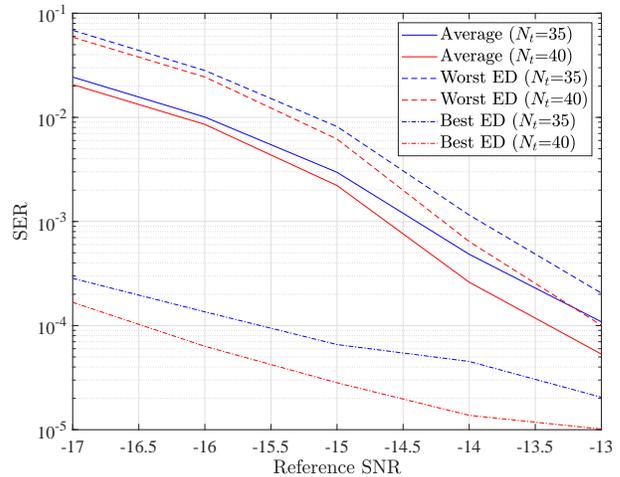}
	\caption{Best, average, and worst SERs versus reference SNR for $N_u=3$ EDs: $N_t = 35$ and $N_t = 40$.}\label{f2}
\end{figure}

Considering the case $N_u=3$, Fig. \ref{f2} plots the SER of the ED with the highest SER among all three EDs (denoted as the worst ED), the SER of the ED with the lowest SER (denoted as the best ED), and the average SER for all three EDs. It can be seen that, for both $N_t = 35$ and $N_t = 40$, the difference between the SER of the worst ED and the average SER is quite small, while the best ED enjoys a far better performance as compared to the average performance.
%%%%%%%%%%%%%%%%%%
%%%%%%%%%%%%%%%%

\begin{figure}[thb!]
	\centering
	\includegraphics[width=0.5\textwidth]{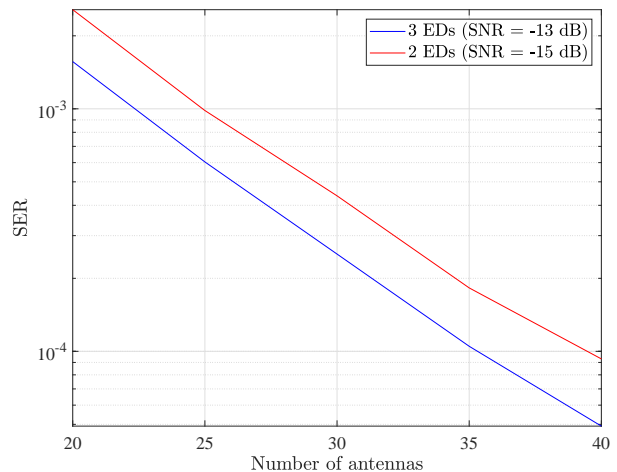}
	\caption{Effect of the number of antennas at each GW for different values of $N_u$ and reference SNR.}\label{f4}
\end{figure}

Fig. \ref{f4} shows the effect of increasing the number of antennas at each GW on the average SER performance. As expected, for both cases of grouping $N_u=2$ and $N_u=3$ EDs in the proposed LoRa network model, increasing the number of antennas can significantly improve the SER performance.
%%%%%%%%%%%%%%%%%%
%%%%%%%%%%%%%%%%

As discussed earlier, the proper selection of $\alpha$ for prioritizing $J_{g,g'}$ over $J_{g,g'}^{\mathrm{(2)}}$ in $\mathcal{OP}_3^{(\kappa)}$ is necessary to balance the detection performance. In particular, as the value of $\alpha$ increases, minimizing $J_{g,g'}$ is prioritized over the minimization of $J_{g,g'}^{\mathrm{(2)}}$.
Figure \ref{f5} shows the effect of $\alpha$ on the SER performance for the proposed LoRa network model with $N_u=3$, $N_t = 40$ antennas, and ${\rm SNR} = - 13$ dB. As can be seen, the SER improves (i.e., decreases) when $\alpha$ increases from 1 to 1.06, which agrees with our observation that the detection error when 2 or more EDs transmitting the same chirp decreases. The SER performance then deteriorates when increasing $\alpha$ beyond 1.06.
%%%%%%%%%%%%%%%%%%
%%%%%%%%%%%%%%%%

\section{Conclusion}\label{sec:con}
In this paper, we have proposed and investigated performance of a novel LoRa network model by allowing multiple EDs to simultaneously transmit information to GWs using the same SF factor and over the same frequency band. By exploiting large antenna arrays at gateways, we developed two-stage sub-optimal detection algorithm to jointly detect information bits of  multiple EDs. The proposed detection algorithm identifies the active frequency bins in the first stage and jointly detects the symbols of EDs in the second stage. A power control policy was also proposed to improve the detection performance in the second stage by minimizing the similarity, measured by the Jaccard coefficient, among expected bin powers between any pairs of EDs. The solution of the power control problem is obtained via successive convex approximation. Simulation results showed the merits of the proposed network model, the importance of power control, and the effectiveness of the two-stage sub-optimal detection algorithm. More importantly, the results demonstrate and justify the tradeoff between transmit power penalties and network scalability of the proposed model. In particular, by allowing concurrent transmission of 2 or 3 EDs, the uplink capacity of the proposed network can be doubled or tripled over that of a conventional LoRa network at the expense of additional 3.0 or 4.7 dB transmit power.

\section*{Acknowledgement}

This work was supported by an NSERC/Cisco Industrial Research Chair in Low-Power Wireless Access for Sensor Networks.

\balance

\bibliographystyle{IEEEtran}

\begin{thebibliography}{10}
\bibitem{figueredo2020preparing}
K.~Figueredo, D.~Seed, and V.~Subotic, ``Preparing for highly scalable and
replicable {IoT} systems,'' \emph{IEEE Internet of Things Magazine}, vol.~3,
no.~3, pp. 94--98, Sep. 2020.

\bibitem{centenaro2016long}
M.~Centenaro, L.~Vangelista, A.~Zanella, and M.~Zorzi, ``Long-range
communications in unlicensed bands: The rising stars in the {IoT} and smart
city scenarios,'' \emph{{IEEE} Trans. Wireless Commun.}, vol.~23, no.~5, pp.
60--67, Oct. 2016.

\bibitem{afisiadis2019error}
O.~Afisiadis, M.~Cotting, A.~Burg, and A.~Balatsoukas-Stimming, ``On the error
rate of the {LoRa} modulation with interference,'' \emph{IEEE Transactions on
	Wireless Communications}, vol.~19, no.~2, pp. 1292--1304, Nov. 2019.

\bibitem{mahmood2018scalability}
A.~Mahmood, E.~Sisinni, L.~Guntupalli, R.~Rond{\'o}n, S.~A. Hassan, and
M.~Gidlund, ``Scalability analysis of a {LoRa} network under imperfect
orthogonality,'' \emph{IEEE Transactions on Industrial Informatics}, vol.~15,
no.~3, pp. 1425--1436, Aug. 2018.

\bibitem{hanif2020US}
M.~Hanif and H.~H. Nguyen, ``Methods for improving flexibility and data rates
of chirp spread spectrum systems in {LoRaWAN},'' Sep. 2020, US Patent
\#10,778,282.

\bibitem{hanif2021frequency}
------, ``Frequency-shift chirp spread spectrum communications with index
modulation,'' \emph{To appear, IEEE Internet of Things Journal}, 2021.

\bibitem{elshabrawy2018closed}
T.~Elshabrawy and J.~Robert, ``Closed-form approximation of {LoRa} modulation
{BER} performance,'' \emph{IEEE Communications Letters}, vol.~22, no.~9, pp.
1778--1781, Sep. 2018.

\bibitem{hanif2020slope}
M.~Hanif and H.~H. Nguyen, ``Slope-shift keying {LoRa}-based modulation,''
\emph{IEEE Internet of Things Journal}, vol.~8, no.~1, pp. 211--221, June
2020.

\bibitem{nguyen2019efficient}
T.~T. Nguyen, H.~H. Nguyen, R.~Barton, and P.~Grossetete, ``Efficient design of
chirp spread spectrum modulation for low-power wide-area networks,''
\emph{IEEE Internet of Things Journal}, vol.~6, no.~6, pp. 9503--9515, Dec.
2019.

\bibitem{baruffa2021}
G.~Baruffa and R.~Luca, ``Performance of {LoRa}-based schemes and quadrature
chirp index modulation,'' \emph{To appear, IEEE Internet of Things Journal},
2021.

\bibitem{nguyen2021performance}
K.~Nguyen, H.~H. Nguyen, and E.~Bedeer, ``Performance improvement of {LoRa}
modulation with signal combining and semi-coherent detection,'' \emph{IEEE
	Communications Letters}, vol.~25, pp. 2889--2893, Sep. 2021.

\bibitem{xu2019discrete}
J.~Xu, P.~Zhang, S.~Zhong, and L.~Huang, ``Discrete particle swarm optimization
based antenna selection for {MIMO LoRa IoT} systems,'' in \emph{Proc.
	Computing, Communications and IoT Applications (ComComAp)}, Oct. 2019, pp.
204--209.

\bibitem{tesfay2020multiuser}
A.~A. Tesfay, E.~P. Simon, I.~Nevat, and L.~Clavier, ``Multiuser detection for
downlink communication in {LoRa}-like networks,'' \emph{IEEE Access}, vol.~8,
pp. 199\,001--199\,015, 2020.

\bibitem{beltramelli2020lora}
L.~Beltramelli, A.~Mahmood, P.~{\"O}sterberg, and M.~Gidlund, ``{LoRa beyond
	ALOHA: An investigation of alternative random access protocols},'' \emph{IEEE
	Trans. Ind. Informat}, vol.~17, no.~5, pp. 3544--3554, Feb. 2020.

\bibitem{luvisotto2018use}
M.~Luvisotto, F.~Tramarin, L.~Vangelista, and S.~Vitturi, ``{On the use of
	LoRaWAN for indoor industrial IoT applications},'' \emph{Wireless
	Communications and Mobile Computing}, 2018.

\bibitem{haxhibeqiri2017lora}
J.~Haxhibeqiri, A.~Karaagac, F.~Van~den Abeele, W.~Joseph, I.~Moerman, and
J.~Hoebeke, ``{LoRa indoor coverage and performance in an industrial
	environment: Case study},'' in \emph{Proc. IEEE International Conference on
	Emerging Technologies and Factory Automation (ETFA)}, Sep. 2017, pp. 1--8.

\bibitem{haxhibeqiri2018low}
J.~Haxhibeqiri, I.~Moerman, and J.~Hoebeke, ``Low overhead scheduling of {LoRa}
transmissions for improved scalability,'' \emph{IEEE Internet of Things
	Journal}, vol.~6, no.~2, pp. 3097--3109, Apr. 2018.

\bibitem{reynders2018improving}
B.~Reynders, Q.~Wang, P.~Tuset-Peiro, X.~Vilajosana, and S.~Pollin, ``Improving
reliability and scalability of {LoRaWANs} through lightweight scheduling,''
\emph{IEEE Internet of Things Journal}, vol.~5, no.~3, pp. 1830--1842, June
2018.

\bibitem{ghanaatian2019lora}
R.~Ghanaatian, O.~Afisiadis, M.~Cotting, and A.~Burg, ``{LoRa} digital receiver
analysis and implementation,'' in \emph{Proc. IEEE International Conference
	on Acoustics Speech and Signal Processing (ICASSP)}, May 2019, pp.
1498--1502.

\bibitem{cha2007comprehensive}
S.-H. Cha, ``Comprehensive survey on distance/similarity measures between
probability density functions,'' \emph{Int. J. Math. Models Methods Appl.
	Sci.}, vol.~1, no.~4, pp. 300--307, Nov. 2007.

\bibitem{Boyd2004convex}
S.~Boyd and L.~Vandenberghe, \emph{Convex Optimization}.\hskip 1em plus 0.5em
minus 0.4em\relax Cambridge University Press, 2004.

\bibitem{grant2010cvx}
%\BIBentryALTinterwordspacing
M.~Grant and S.~Boyd, \emph{{CVX}: Matlab Software for Disciplined Convex
	Programming}, 2010. [Online]. Available: \url{cvxr.com/cvx.}
%\BIBentrySTDinterwordspacing

\bibitem{nguyen2020max}
T.~K. Nguyen, H.~H. Nguyen, and H.~D. Tuan, ``Max-min qos power control in
generalized cell-free massive {MIMO}-{NOMA} with optimal backhaul
combining,'' \emph{IEEE Transactions on Vehicular Technology}, vol.~69,
no.~10, pp. 10\,949--10\,964, Oct. 2020.

\bibitem{petajajarvi2015coverage}
J.~Petajajarvi, K.~Mikhaylov, A.~Roivainen, T.~Hanninen, and M.~Pettissalo,
``On the coverage of {LPWANs}: range evaluation and channel attenuation model
for {LoRa} technology,'' in \emph{Proc. IEEE International Conference on
	{ITS} Telecommunications (ITST)}, Dec. 2015, pp. 55--59.	
\end{thebibliography}

\end{document}